\documentclass[12pt]{iopart}

\usepackage{graphicx}   
\begin{document}
\title{The Paradox of Power Loss in a Lossless Infinite Transmission Line}
\author{Ashok K. Singal}
\address{Astronomy and Astrophysics Division, Physical Research Laboratory,
Navrangpura, Ahmedabad - 380 009, India }
\ead{asingal@prl.res.in}
\vspace{10pt}
\begin{indented}
\item[]October 2015
\end{indented}
\begin{abstract}
We discuss here the famous paradox of a continuous power drainage from the source at the input of an otherwise lossless infinite 
transmission line. The solution of the paradox lies in the realization 
that in an open-circuit finite transmission line/ladder network, there is an incident as well as a reflected wave 
and the input impedence is determined by the superposition of both waves.  
It is explicitly shown that the reactive input impedance of even a single block, comprising 
say a simple LC circuit, is determined at all driving frequencies from the superposition of incident and reflected waves, 
and that the input impedance remains reactive in nature (i.e., an imaginary value) even when additional blocks are added indefinitely.
However in a ladder network or transmission line, taken to be {\em infinite right from the beginning}, 
there is no reflected wave (assuming the circuit to be ideal with no discontinuities en route). Thus the source while 
continuously supplying power in the forward direction, does not retrieve it from a reflected wave and unlike in the case of a finite line, 
there is a net power loss. This apparently lost energy ultimately appears in the electromagnetic fields 
in the reactive elements (capacitances and inductances which to begin with had no such stored energy), further down 
the line as the incident wave advances forward. It is also shown that radiation plays absolutely no role in resolving this intriguing 
paradox. 
\end{abstract}
\section{Introduction}
A transmission line is a channel for transmitting electric signals or power from one point to another 
along a guided path \cite{1,2,3,4}. A line could be of a finite length or be of infinite length (at least in principle).
A circuit comprising lumped parameters is generally called a ladder network, on the other hand if it consists of a continuous distribution 
of parameters, then it is usually called a transmission line. The two are almost identical in their behavior \cite{1}. 
The elements of a transmission line could be either reactances (with no power dissipation within them) like capacitances or 
inductances, or could comprise resistances or shunt leakage conductances, which dissipate power into heat. Most lines will have a 
mixture of reactances and dissipative elements. An ideal transmission line may be thought of as the one which delivers signal or power 
across its length without any dissipation on the way. Intuitively one would think a line devoid of elements like resistances  
should behave as a lossless line without a continuous power drainage from the source at the input, 
and this does seem to hold true for a line of finite length. However, for an infinite line, even if there were no resistive elements 
along its length that could dissipate power, the line presents a {\em real value} of 
input impedance, implying that power will be drained from the source at a constant rate \cite{11}.

Where does this energy go as it is not dissipated in the inductors and 
capacitors of the circuit? For this Feynman \cite{4} writes ``But how can the circuit continuously absorb energy, as a resistance 
does, if it is made only of inductances and capacitances? {\em Answer}: Because there are an infinite number of inductances 
and capacitances, so that when a source is connected to the circuit, it supplies energy 
to the first inductance and capacitance, then to the second, to the third, and so on. 
In a circuit of this kind, energy is continually absorbed from the generator at a constant 
rate and flows constantly out into the network, supplying energy which is stored in the 
inductances and capacitances down the line.''

In an alterntive approach \cite{5,12} it has been shown that the input impedance of an open-circuit ladder network, initially consisting of 
a finite number of blocks comprising inductors and capacitors, does not converge to a unique fixed value when additional identical 
blocks are added, and always yields pure imaginary (reactive) input impedance value irrespective of the number of the blocks added. 
The input impedance does not have a real (dissipative) part for any driving frequency, even when the number of blocks is increased 
indefinitely. This contradicts Feynman's observation \cite{4} that the infinite 
ladder network has an input impedance which has a real part at frequencies below certain value. It was argued afterwards \cite{6} 
that a non-zero real part of impedance appears only if there is a termination in an impedance that has a real part 
and that a circuit consisting solely of components with purely imaginary impedances has a purely imaginary input impedance.
Later the behavior of infinite ladder network, its convergence and solutions have been analyzed in a greater detail \cite{7,8}. 

In this paper we examine this intriguing paradox from a fresh view point trying to understand why two alternate approaches lead to conflicting 
results. We will first review the relevant characteristics of a transmission line/ladder network; the detailed description of various terms and 
the derivation of the formulas used can be found in standard text-books \cite{1,2,3,4}. Then we shall show how one arrives at a paradoxical result 
of an uninterrupted power drain in an otherwise lossless infinite transmission line. This will be followed by a brief account of the alternative 
approach of extending a finite ladder network by the addition of further blocks, with the circuit always comprising only reactive elements. 
Subsequently we shall present the resolution of the paradoxical results both for a ladder network as well as the transmission line; the 
resolution basically ensues the realization that there is an absence of a reflected wave in an infinite ladder network or a 
transmission line. Reflection plays a role in resolving the paradox was briefly mentioned in \cite{8} but without much further elaboration,  
which we do here in detail by explicitly calculating the input impedance of a finite ladder network by a superposition of incident 
and reflected waves. We shall demonstrate that unlike in a finite case, where a termination in a load matched to the characteristic impedance 
of the line could dissipate all power, or at an open-ended termination could reflect it all back towards the source, in the case of an 
infinite line there is no termination point to start a reflected wave (assuming of course no discontinuities along the line to 
trigger any reflection) and that results in the current being in phase with the voltage and power being drawn from the source. 
\section{A non-ideal behavior of an ideal circuit}
\subsection{A transmission line}
A transmission line is described by its line parameters $R, L, C, G$, where $R$ is the series resistance per unit length of line 
(including both wires), $L$ is the series inductance per unit length of line, $C$ is the capacitance between the two conducing 
wires per unit length of line and $G$ is the shunt leakage conductance between the two conducing wires per unit length of line. 
For an incremental length $\Delta z$ of the line, the equivalent circuit is shown in Fig. 1.
The increments in voltage and current along the line are \cite{1,2,3,11},
\begin{eqnarray}
\label{1}
\Delta V(z) = -I(z)(R + j\omega L) \Delta z\\ 
\label{2}
\Delta I(z) = -V(z)(G+j\omega C) \Delta z.
\end{eqnarray}
These could be written in limit $\Delta z\rightarrow0$ as, 
\begin{eqnarray}
\label{3}
{\rm d}V(z)/{\rm d}z = -I(z)(R + j\omega L) \\ 
\label{4}
{\rm d}I(z)/{\rm d}z = -V(z)(G+j\omega C).
\end{eqnarray}

From Eqs. (\ref {3}) and (\ref {4}) one gets a general solution for voltage along the line,
\begin{eqnarray}
\label{25}
V(z) = V'_0 e^{-\gamma z}+ V''_0 e^{\gamma z}\\
\label{8}
\gamma = \sqrt{\left(R+j\omega L\right)\left(G+j\omega C\right)} = \alpha + j\beta,
\end{eqnarray}
where $\gamma$ is the propagation constant. The phasor part is written with an assumed $e^{j \omega t}$ time dependence throughout.
Now of the two terms in Eq. (\ref {25}), the first one represents a wave traveling along increasing $z$ commencing at $z=0$, while the second 
represents a wave traveling towards decreasing $z$ which in case of an infinite line would have to start from $z=\infty$ an 
infinite time back and thus must be dropped. Therefore the voltage along an infinite transmission line can be written as, 
\begin{eqnarray}
\label{5}
V(z) = V_0\: e^{-\gamma z} = V_0\: e^{-\alpha z} e^{-j\beta z}. 
\end{eqnarray}
\begin{figure}[h]
\scalebox{0.6}{\includegraphics{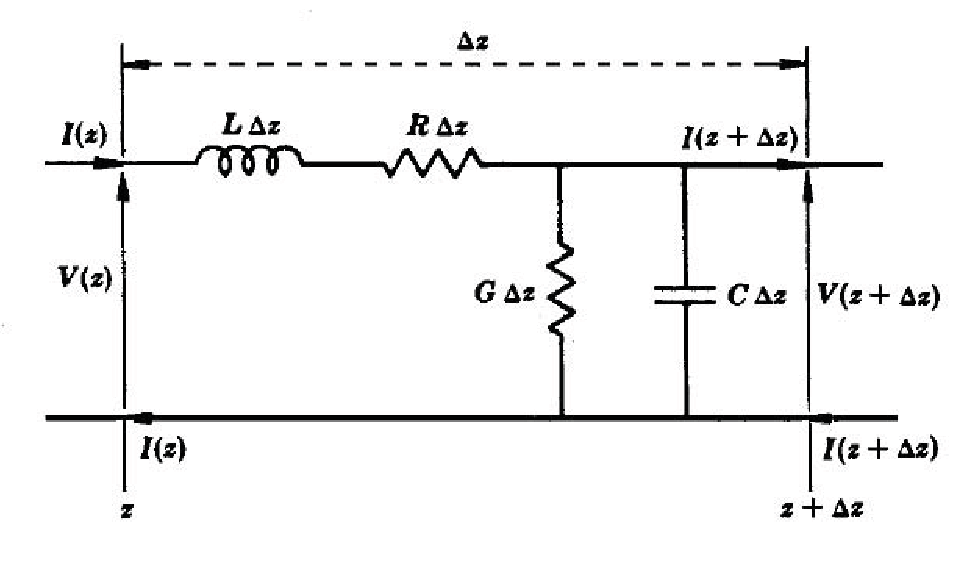}}
\caption{Increments of voltage and current over an incremental length $\Delta z$ of the transmission line.}
\end{figure}
From this one gets for the electric current,  
\begin{eqnarray}
\label{6}
I(z) = \left(V_0/Z_0\right) e^{-\gamma z} = \left(V_0/Z_0\right) e^{-\alpha z} e^{-j\beta z}.
\end{eqnarray}
Here $Z_0$, the characteristic impedance of the line given by, 
\begin{eqnarray}
\label{7}
Z_0 = \sqrt{\left(R + j\omega L\right)/\left(G+j\omega C\right)}.
\end{eqnarray}
Equations (\ref{5}) and  (\ref{6}) represent an attenuated sinusoidal wave along $z$, with $\alpha$ as the attenuation constant and 
$\beta=2\pi/\lambda$ as the wave number. 

For an infinite line, the input impedance (at $z=0$) is calculated from Eqs. (\ref{5}) and (\ref{6}) as,
\begin{eqnarray}
\label{9}
Z_i = V(0)/I(0) = Z_0.
\end{eqnarray}
In a lossless line, R = 0 and G = 0, and from Eqs. (\ref{8}) we have, $\alpha = 0$ and $\beta = \omega \sqrt{LC}$, i.e., 
a sinusoidal wave without any attenuation along the line. But we also 
have $Z_{i} = Z_0 = \sqrt{L/C}$, i.e., its impedance has a {\em real} value. This is a paradox 
because though the transmission line contains no resistive element so there could be no Ohmic losses in the line, 
yet its input impedance is a pure resistance. That means for an input voltage $V_0$, 
power will be drained from the source at the rate of V$^2_0/(2\sqrt{L/C})$ \cite{11}. 
The questions therefore arise as to why does a pure resistance show up in a circuit 
comprising only reactances, thereby implying a continuous power drainage and where does this energy ultimately go? 

The paradox can be also seen from the Smith chart where 
the input impedance of a lossless open-circuit line, goes through cycles when its length is varied. Not only does the 
input impedance not converge to a single unique value when the length of the line is increased indefinitely but 
also in general it is an {\em imaginary} value, i.e., a pure reactance \cite {1,2,3} for any length of the line, 
which contradicts the conclusion that the infinite line presents a {\em real} input impedance. 
\subsection{A ladder network}
\begin{figure}[ht]
\scalebox{0.5}{\includegraphics{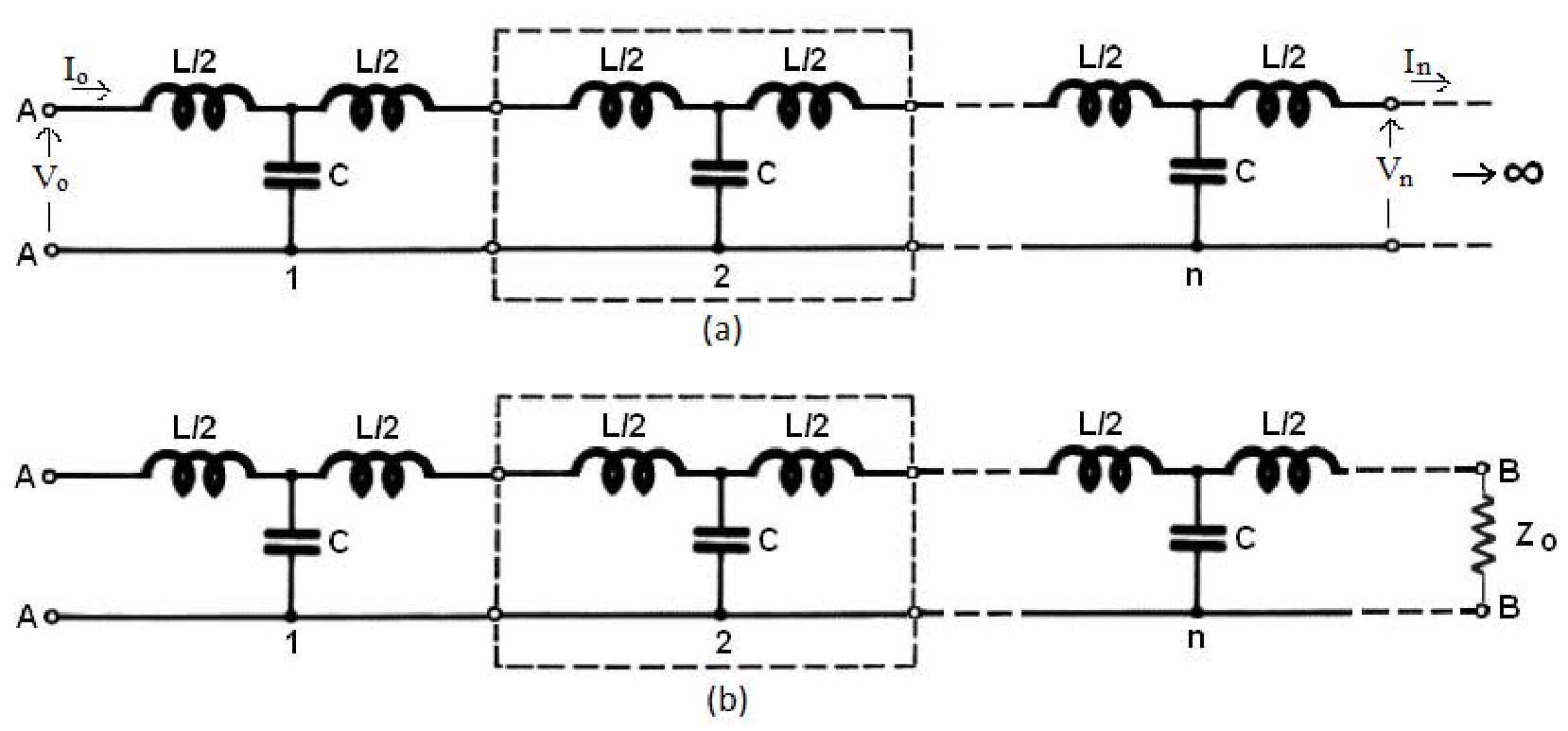}}
\caption{(a) An infinite ladder network comprising lumped parameters. 
(b) A finite ladder network terminated in its characteristic impedance $Z_0$.  A typical block in the network 
is shown by a rectangular box of dashed lines.}
\end{figure}
A transmission line with distributed parameters is almost identical in behavior to a ladder network comprising lumped 
parameters \cite{1}, and the above paradox appears in the infinite ladder network too.
A Ladder network of $n$ blocks, with each block a symmetrical $T$ section consisting of two $L/2$ inductances 
and a capacitance $C$, has a characteristic impedance $Z_0=\sqrt{L/C - \omega^2 L^2/4}$) \cite{1,2,3}.
The number $n$ of blocks could be finite, or it could even be infinite 
($n\rightarrow \infty$). Figure 2(a) shows an infinite ladder network while Fig. 2(b) shows a finite ladder network, but 
terminated in its characteristic impedance $Z_0=\sqrt{L/C - \omega^2 L^2/4}$). 

A solution for the input impedance $Z_i$ of the infinite network is obtained in the following manner \cite{4,5,12,9,10}.
Since adding another block to the beginning of an infinite ladder network does not 
change the input impedance (it still remains the same infinite network), $Z_i$ must equal the 
impedance of a circuit having a single block terminated in a load impedance equal to $Z_i$. 
Therefore we have,
\begin{eqnarray}
\label{10}
Z_i = \frac {j\omega L}{2} + \frac {(Z_i+j\omega L/2)(1/j\omega C)}{Z_i+j\omega L/2+1/j\omega C},
\end{eqnarray}
which has a solution,
\begin{eqnarray}
\label{11}
Z_i=\sqrt{L/C - \omega^2 L^2/4}.
\end{eqnarray}

The input impedance of the infinite network equals its characteristic impedance, i.e., $Z_i=Z_0$, and the circuit behaves 
as if it were terminated in $Z_0$ somewhere along the line as in Fig. 2(b). 
Now for $\omega < \omega_0 = 2/\sqrt{LC}$, $Z_i$ is a {\em real} value. This leads to the same paradox as for the infinite 
transmission line of distributed parameters -- how come a circuit containing only purely imaginary impedances has for its input 
impedance a {\em real} value which could absorb energy continuously? 
\section{Where does the energy disappear? -- Could radiation losses be the answer?}
Could the energy be lost into the surrounding medium 
by the process of radiation, with $Z_0=\sqrt{L/C}$ as the radiation resistance? 
In transmission line or ladder network containing resistive elements, power loss by the source is fully accounted for 
by the energy dissipation in the circuit, for any value of $R$ and $G$. 

Consider the lossy infinite line (i.e., with $R$ and $G$ non-zero), where
input power from the source is \cite {11},
\begin{eqnarray}
\label{26}
P_i = [V^2_0/(2|Z_0|)] \cos (\angle{Z_0})
= [V{^2_0}/(2|Z_0|^2)]Re(Z_0).
\end{eqnarray}
On the other hand the power dissipated in an infinitesimal line element (Fig. 1) is, 
\begin{eqnarray}
\label{27}
{\rm d}P_d  
 = (1/2)\left(|I(z)|^2R+|V(z)|^2G\right){\rm d}z\\
\label{27a}
= (V^2_0/2)\left[\left(R/|Z_0|^2\right) +G\right]e^{-2\alpha z}{\rm d}z
\end{eqnarray}
Hence the total power dissipated in the infinite line is
\begin{eqnarray}
\label{28}
P_d 
= (V^2_0/2)\left[\left(R/|Z_0|^2\right)+G\right] \int_{0}^{\infty}e^{-2\alpha z} {\rm d}z\\
\label{28a}
=(V^2_0/2)\left[\left(R/|Z_0|^2\right)+G\right]/\left(2\alpha\right)
\end{eqnarray}
Substitution for $|Z_0|$ and $\alpha$ shows that
$P_d = P_i$ \cite{11}, and all power losses are accounted for without anything going into radiation. This is true for all $R$ and $G$, 
in particular  even when in limit $R\rightarrow0$ and 
$G\rightarrow0$.  Now it cannot happen that when $R=0$ and $G=0$ radiation suddenly shows up into picture from somewhere. 
Further, even in a lossless line, all the power (assumed to be lost by the source) can at any stage be either reflected back by 
making the circuit open just after that point, or it could be consumed 
by terminating the line in its characteristic impedance, irrespective of the length of the line up to that stage. 
This implies that up to any {\em arbitrarily selected length} of the line, the radiation losses had not yet taken place. 
Therefore for resolving this paradox there does not seem any scope for radiation hypothesis at all and 
a satisfactory resolution of the paradox lies elsewhere.
\section{The paradox reappears!}
Actually while writing Eq. (\ref{10}) for $Z_i$ one 
implicitly assumed that the infinite series {\em converges} to a unique value and it is only under this existence supposition 
that a unique solution Eq. (\ref{11}) could be obtained. If the series does not converge, then of course this basic assumption itself 
breaks down and the solution obtained thereby may not represent a true value. 

On the other hand if one started with an open-circuit ladder network of a finite number of identical blocks comprising inductors and capacitors, 
and then added more similar blocks, the input impedance does not converge to a unique fixed value even when the number of blocks is increased 
indefinitely. Moreover, the input impedance always turns out to be a pure imaginary value with no  
real (dissipative) part for any driving frequency, even when the number of blocks approaches infinity. 

It seems that the infinite ladder networks of type in Fig 2(a) may have different answers for the input 
impedance, and thereby implying different power consumptions depending upon the method of solution. 
Hence a paradox still exists as one arrives at different answers using different arguments, and a question still remains whether 
or not does an infinite ladder network converge to a pure resistance drawing continuous 
power from an input source, and if so where does this energy go. What could be the missing factor, if any, in these arguments?
\section{Resolution of the paradox -- incident versus reflected waves}
Here we demonstrate with a detailed analysis that the resolution of the paradox lies in the realization that there is an absence of a reflected wave 
in an infinite ladder network or an infinite transmission line. Then we shall also understand why the two alternate approaches led to two 
conflicing conclusions. 
\subsection{The case of an infinite ladder network}
\subsubsection{The ladder network at low frequencies}

For frequencies below a critical value $\omega_0 = 2/\sqrt{LC}$, the characteristic impedance $Z_0=\sqrt{L/C - \omega^2 L^2/4}$ 
can be written as $Z_0 =(\omega L/2) \sqrt{(\omega_0/\omega)^2-1}$, which is a {\em real} quantity, meaning a pure resistance. Let us 
consider the propagation factor $e^{-\gamma}$ between adjacent blocks calculated by terminating the ladder network in $Z_0$. 
This is only to ensure that there is no reflected wave and thus one is dealing only with the incident wave. 
In the low frequency ($\omega < \omega_0$) case one gets \cite{1,4},
\begin{eqnarray}
\label{12}
\frac{V'_{n}}{V'_{n-1}}=\frac{I'_{n}}{I'_{n-1}}=\frac{\sqrt{L/C - \omega^2 L^2/4}-j\omega L/2}{\sqrt{L/C - \omega^2 L^2/4}+j\omega L/2}.
\end{eqnarray}
We can simplifying the Eq. (\ref{12}) to get, 
\begin{eqnarray}
\label{13.1}
\frac{V'_{n}}{V'_{n-1}}=1-\omega^2 LC/2-j\sqrt{\omega^2LC}\sqrt{1 - \omega^2 LC/4}\\
\label{13}
=1-2(\omega/\omega_0)^2 -j2(\omega/\omega_0)\sqrt{1 - (\omega/\omega_0)^2}.
\end{eqnarray}
A prime ($'$) over voltages and currents merely indicates that these represent an incident wave.
From the real and imaginary parts in Eq. (\ref {13}), it can be readily seen that the propagation factor has a unit magnitude 
and represents a simple phase change $e^{-j\beta}=\cos \beta-j\sin \beta$, between successive blocks in the network. 

Although for calculating the propagation factor of the circuit we needed to isolate the {\em incident wave} by terminating this network 
with its characteristic impedance $Z_0$, yet the propagation properties of the incident wave (that is, the propagation constant 
calculated from Eq. (\ref{13}) of {\em incident} wave between two neighboring blocks, say, $n-1$ and $n$) does not depend 
upon this termination. The incident wave has an input impedance everywhere equal to the characteristic impedance $Z_0$ of the 
network. Of course the voltages and currents at any point are decided by the superposition of the incident and reflected waves at 
that point and the input impedance of a network as calculated in \cite{5,7,8,9} is actually what results 
from the superposition of the incident and the reflected wave with their phases duly taken into account. 

To prove our assertion that this indeed is the case in general, we want to calculate input impedance of an  
open-circuit line, made of any finite number of blocks (say, $n$), by evaluating voltage and current at $z=0$ due to the sum of the 
incident and reflected waves, the latter arising from the termination just after the $n$th block. 
For a cascaded network of $n$ identical blocks, 
the propagation factor is simply $e^{-jn\beta}$. The angle $\beta$ here is half of $\theta$ defined in Eq. (21) 
of that in \cite{7}. If the network has a total of $n$ blocks, then voltage $V_0$ at $z=0$ includes a reflected wave with a 
phase  change of angle $2n\beta$ from the incident wave, while the current $I_0$ has a phase change of angle $2n\beta+\pi$ 
(an extra phase of angle $\pi$ in the current wave at the refection point). Therefore the input impedance is given by,
\begin{eqnarray}
\label{14}
Z_i= V_0/I_0 = Z_0\frac{1+e^{-j2n\beta}}{1-e^{-j2n\beta}}=-j Z_0\cot(n\beta).
\end{eqnarray}
We see that the calculated input impedance is the same what was calculated in an alternative method for a finite open-circuit 
ladder network \cite{7,8}, which thus proves our assertion that the propagation factor of the incident wave is unaffected by 
the termination impedance. As $n$ is increased, $Z_i$ is always of an {\em imaginary} value which goes through cycles, 
even becoming $0$ or $\infty$, and in general does not converge to a unique value even when $n\rightarrow \infty$.
\subsubsection{Energy transport - a physical perspective}
In a finite open-circuit network there is a reflected wave from its terminated end as it has to match the 
conditions for a zero net current (implying the electric currents out of phase by angle $\pi$ for the incident and the reflected 
waves), although the voltages will be in phase for the incident and reflected waves at the termination point. 
It is important to note that when we analyze a {\em finite} network, barring transients, 
the voltages and currents being considered are the superposition of incident and reflected 
waves. Therefore the calculated $Z_i$ may depend upon the length of 
the line or equivalently the number of blocks in the network as that would determine the relative phases of the incident and 
reflected waves at the input point. 

Suppose a generator is connected to the circuit at input terminals AA (Fig. 2(a)).  The generator drives the circuit at a frequency 
$\omega$ (say) and will give rise to a voltage as well as a current in the 1st electric block, which (a pure reactance) does not consume 
the electric power itself, and in turn gives rise to voltage and current in the 2nd block and
so on. As we showed above, for $\omega < \omega_0$, there will be no decrease in the amplitude of voltage or current from 
one block to the next and there will only be a progressive phase change from one block to the next. This ``incident wave'' will move 
along the network until a discontinuity, say an open-circuit termination after $n$th block, is encountered which will cause a reflection 
wave towards the block $n-1$, then $n-2$ and so on towards the generator. 

The generator meanwhile will keep on supplying further power to the 1st block which gets passed forward, till it is finally reflects 
back towards the generator. This is true even when the line terminates just after block one (just a simple LC circuit, see Appendix). 
And when we add more blocks, 
then the discussion still entails reflected waves implicitly. However when we consider an infinite ladder network or an infinite 
transmission line, all by itself (and not by an indefinite extension of finite network by adding more blocks or increasing length 
of the line), then we do not consider the reflected wave since the incident wave does not ever reach the termination point to 
start a reflected wave. 

In that case we have only the incident wave and the source at the input 
keeps on continuously supplying power to the network or line but does not get it back through reflected wave. 
Therefore in an infinite network, 
it results in a net power drain from the source and this energy of course appears from one block to the next down the line 
where it has not yet reached due to the long extent of the line. Of course as it will never reach a termination point (at infinity!), 
so the energy transfer to further blocks continues for ever.
Initially none of the blocks had electric energy (say just before time $t=0$ when the generator 
was just connected), but afterwards up to a certain stage the blocks have stored electric energy (shared between its capacitor 
and inductor or equivalently between 
the electric and magnetic fields and getting continuously exchanged between them). Ultimately this energy has come from the 
generator. The energy is not lost as it can be still consumed by terminating the circuit in a matched load somewhere down the line 
or recovered by terminating the line as an open circuit and getting the energy returned as a reflected wave. 

\subsubsection{The ladder network at high frequencies}
The propagation factor $e^{-\gamma}$ between adjacent blocks, for a high frequency ($\omega > \omega_0$) case can be written as,  
\begin{eqnarray}
\label{21}
\frac{V'_{n}}{V'_{n-1}}=1-2(\omega/\omega_0)^2 + 2(\omega/\omega_0)^2\sqrt{1-(\omega_0/\omega)^2}.
\end{eqnarray}
From Eq. (\ref{21}) it can be seen that for $\omega > \omega_0$ the propagation factor, written as 
$e^{-(\alpha+j \pi)}=-(\cosh \alpha-\sinh \alpha)$, is of magnitude less than unity and 
is always of a negative value, implying a phase change of angle $\pi$ between successive blocks accompanied by an 
exponential decrease in amplitude. 
The voltages and currents do not penetrate too far in the circuit, and there is no continuous transport of energy along $z$. 
The input impedance at frequencies $\omega > \omega_0$ for a cascade network of $n$ blocks is,
\begin{eqnarray}
\label{22}
Z_i= V_0/I_0 = Z_0\frac{1+e^{-2n\alpha}}{1-e^{-2n\alpha}}= Z_0\coth(n\alpha),
\end{eqnarray}
which is imaginary, in spite of $\coth(n\alpha)$ being always a real value. This is because the characteristic impedance 
$Z_0 = (j\omega L/2)\sqrt{1-(\omega_0/\omega)^2}$ is imaginary for $\omega > \omega_0$. 
While the incident wave alone presents a {\em real} input impedance for frequencies 
$\omega < \omega_0 = 2/\sqrt{LC}$, when a superposition of 
incident and reflected waves is considered then we get an {\em imaginary} input impedance for all driving frequencies. 
That means the current at the input will be $\pi/2$ out of phase 
with the voltage, so that there will be no continuous power being absorbed from the source.  
For $n \rightarrow \infty$, $Z_i\rightarrow Z_0$, a pure reactance, thus there is no paradox for the $\omega > \omega_0$ case. 

\subsection{Infinite transmission line}
The characteristic impedance of a ladder network in case of a lossless ideal transmission line with distributed parameters (Fig. 1) 
could be written as $Z_0=\sqrt{(L\Delta z)/(C\Delta z) - \omega^2 L^2(\Delta z)^2/4}$ which reduces to $Z_0 = \sqrt{L/C}$ as 
in limit $\Delta z\rightarrow 0$. Therefore unlike the ladder network case, in the transmission line case there is no cutoff 
frequency and for {\em all} driving frequencies a wave travels along the line  without any amplitude attenuation
since propagation constant has an {\em imaginary} value implying only a phase change. 

In general the input impedance of a line of length $l$ is given by \cite{1}, 
\begin{eqnarray}
\label{16}
Z_i = Z_0\left(\frac{e^{\gamma l}+Ke^{-\gamma l}}{e^{\gamma l}-Ke^{-\gamma l}}\right),
\end{eqnarray}
where $K (=$reflected voltage at load/incident voltage at load) is the reflection coefficient, 
\begin{eqnarray}
\label{17}
K = \left(\frac{Z_r-Z_0}{Z_r+Z_0}\right).
\end{eqnarray}
where $Z_r$ is the impedance at the receiving (load) end. The input impedance reduces to $Z_0$ when there is no reflected wave, 
i.e., when $K=0$. 

Now the absence of a reflected wave in a transmission line can be due to three reasons. First, the line is finite but terminates 
in a load matched to the characteristic impedance of the line, i.e., when $Z_r=Z_0$. Second, the line has small resistance 
which can causes the incident voltage to die over its long length $l$, i.e., if $\gamma l \rightarrow \infty$, 
so that the amplitude of the 
incident and thence of the reflected wave is zero, then the series does converge to a unique solution \cite{5,9} which is consistent 
with $Z_i=Z_0$. Thirdly the line is lossless but truly of infinite extent so that it could be assumed that 
the incident wave, which assumedly started a finite time back, has not yet reached the termination point to start a reflected wave. 
In all three cases, the input impedance, 
which is the ratio of the voltage and current at the input point, is the same as that is not affected by what happens 
at its termination point, and we obtain the same result for the input impedance, viz. $Z_i=Z_0$.

On the other hand, for an open-circuit line of finite length $l$ ($Z_r=\infty$, $K=1$), the input impedance is given by, 
\begin{eqnarray}
\label{18}
Z_i = Z_0\left(\frac{e^{\gamma l}+e^{-\gamma l}}{e^{\gamma l}-e^{-\gamma l}}\right).
\end{eqnarray}
In a lossless line, $\gamma = j\beta = j\omega \sqrt{LC}$, the input impedance becomes,
\begin{eqnarray}
\label{19}
Z_i& = &Z_0\left(\frac{e^{j\beta l}+e^{-j\beta l}}{e^{j\beta l}-e^{-j\beta l}}\right)\\
\label{20}
& = & -j Z_0 \cot (\beta l)=-j Z_0 \cot (2\pi l/\lambda),
\end{eqnarray}
which is a pure reactance, and thereby no net power consumed, and which is similar to the result derived for the ladder 
network Eq. (\ref {14}). It should be noted that in case of a ladder network, the quantities $\gamma, \alpha, \beta$, or even $L, C$ etc. 
are specified as per block of the circuit while in the case of a transmission line with distributed parameters all such quantities 
are defined per unit length of the line. Therefore in Eq. (\ref {14}) it is the phase angle change $n\beta$ over $n$ blocks while in 
Eq. (\ref{20}) it is the phase angle change $\beta l$ over length $l$ of the line.
In fact with increasing $l$, $Z_i/Z_0$ from Eq. (\ref{20}) is cyclic and is indeed the value read from the Smith chart. 
One thing that we notice from Eq. (\ref{20}) is that the input impedance $Z_i$ depends on the length $l$ of the line in terms 
of wavelength $\lambda$. Thus depending upon $2\pi l/\lambda$, $Z_i$ could be zero, a finite value or even infinity, but always 
a pure {\em imaginary} value, with a zero {\em real} part similar to what was seen for the ladder network in IV.A.{\em 1}.
Here as much amount of power is reflected back to the generator as much it supplies in the incident wave.

In the case where there is only an incident wave, i.e., there is no reflected wave, the current is in fact in phase with the voltage, 
implying power is being drawn from the source. 
However, if there is a reflected wave as well, then the voltage and current are not in phase everywhere. 
Thus it is the absence of reflected wave in infinite transmission line that results in a continuous positive energy flux along 
the line. The relative phases of $V$ and $I$ depend upon the reflected wave, which in 
turn depends upon at how far away along the line reflection took place. Of course 
no reflection will ever take place in a uniform infinite line as the incident wave will never reach the termination point 
which is at infinity. However if we consider the lossless case when there is a reflected wave from an open-circuit termination, 
then equal power is being returned 
to the source by the reflected wave and in that case the current is indeed $\pi/2$ out of phase with the voltage (Eqs. (\ref{14}) and
(\ref{20}).

If we consider a transmission line with no discontinuities whatsoever, then it will have to be an infinite line and the  
energy will be getting stored as electric and magnetic fields in its reactive elements further and further along the line. 
There is no violation of the energy conservation, and since there is no reflected wave to restore energy to the source, the latter 
would be continuously supplying energy, which  gets stored in electric and magnetic fields in more and more inductances and capacitances 
down the line. Seen this way there does not seem to be any paradox.

The paradox actually had arisen only because we were comparing two sets of solutions which are for quite different situations. 
One involves only an incident wave (i.e., without any reflected wave) and then the input impedance $Z_i=Z_0$ is a {\em real} 
quantity, and the voltages and currents are in phase everywhere along the circuit, with energy getting apparently ``spent'' as 
it is getting stored in the inductors and 
capacitors down the line as the incident keeps on advancing for ever in an infinite transmission line. 
The other solution was for the case with a reflected wave, and there the superposition 
of the incident and reflected waves results in $Z_i$ 
to have {\em imaginary} value with no {\em net} power loss since the source gets the energy back as the reflected wave.
\section{Conclusions}
It was shown that while an open-circuit finite ladder network or a transmission line with distributed network  has 
a characteristic impedance $z_0$ which is only reactive (imaginary), an infinite ladder network or 
an infinite transmission line has a finite {\em real} component of the input impedance.
It was shown that the famous paradox of power loss in a lossless infinite transmission line is successfully resolved 
when one takes into account both the incident and reflected waves. The solution of the paradox lies in the realization 
that there is an absence of a reflected wave in an infinite transmission line. In a finite transmission line or ladder network, 
the source still keeps on supplying power as an incident wave but gets it equally back in terms of the reflected wave. 
Therefore there is no further net power transfer from the source which is consistent with the reactive elements  
presenting zero net resistance. 

However in the case of an infinite ladder network or an infinite transmission line 
there is no discontinuity to start a reflected wave, thus the source supplies power in a forward direction, but does not 
get it back in terms of a reflected wave from the termination point. Therefore there is an apparent net power loss, 
which actually appears as stored energy in its reactive elements (capacitances and inductances) further down the line.
It was also shown that radiation plays absolutely no role in resolving this paradox. 

\section{Appendix \\ Input impedance of a driven LC circuit computed from a superposition of incident and reflected waves}
Here we explicitly demonstrate that a driven LC circuit can be treated as an open-circuit 1-block ladder network  
having incident and reflected waves and from their superposition, the voltages and currents, and in particular, input impedance 
of the LC circuit can be calculated for all driving frequencies. 
We denote by $V_0,I_0$ and $V_1,I_1$ the voltages and currents at the input (AA) and termination (BB) respectively, and which (Fig. 3(a)) 
are related by $V_0-V_1=j\omega L I_0/2$ , $I_0=j\omega C V_1$ , where $\omega$ is the frequency at which the circuit is 
being driven by, say, a generator at the input end AA. The input impedance $Z_i=V_0/I_0$ is given by,  
\begin{eqnarray}
\label{30}
Z_i=j\omega L/2 +1/(j\omega C).
\end{eqnarray}
Denoting voltages and currents for the incident and reflected waves by $V',I'$ and $V'',I''$ respectively, 
the boundary conditions at open end BB in Fig. (3a) imply $V''_{1}=V'_{1}$ and $I''_{1}=-I'_{1}$, the minus sign arising because the 
reflected current is out of phase with the incident wave by an angle $\pi$, so as to make the net current $I_1=I'_{1}+I''_{1}=0$. 
However to evaluate $I'_1$, we need to isolate the incident wave and which can be done by terminating the circuit in its characteristic 
impedance $Z_0$ (Fig. 3(b)). The propagation factor for the incident wave from Eq. (\ref{13.1}) is,
\begin{figure}[h]
\scalebox{0.5}{\includegraphics{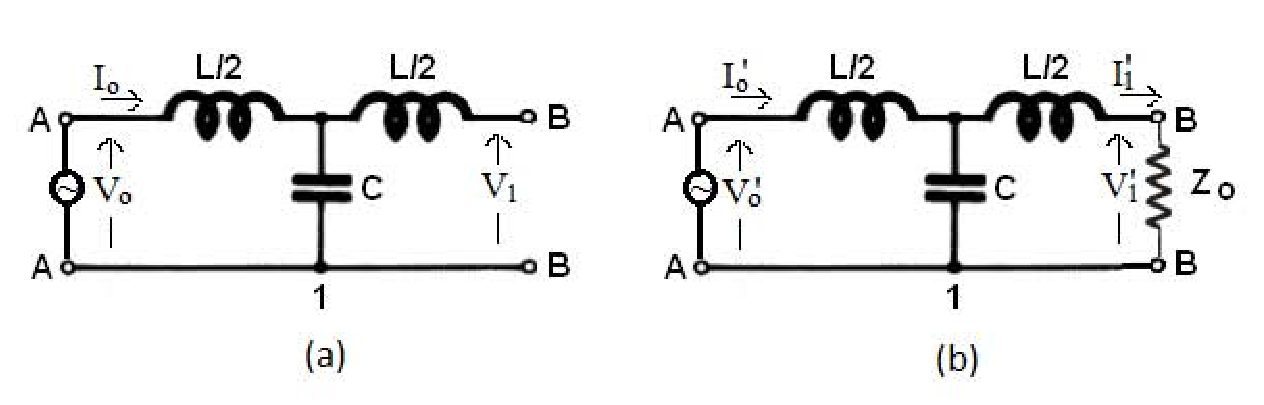}}
\caption{A driven LC Circuit or a single-block network (a) open-circuited (b) terminated in its characteristic impedance $Z_0$ and thereby 
carrying only the incident voltage and current with no reflection at the end BB.}
\end{figure}
\begin{eqnarray}
\label{31}
\frac{V'_{1}}{V'_{0}}=\frac{I'_{1}}{I'_{0}}=1-\omega^2 LC/2-j\sqrt{\omega^2LC}\sqrt{1 - \omega^2 LC/4},
\end{eqnarray}
with $V'_{0}/{I'_{0}}=V'_{1}/{I'_{1}}=Z_0$. As demonstrated in IV.A.{\em1}, incident wave is not attenuated, 
irrespective of the termination impedance. The only difference is that there is also a reflected wave in the open circuit case (Fig. 3(a)), 
while there is no reflected wave when the circuit is terminated in its characteristic impedance $Z_0$ (Fig. 3(b)).

For the reflected wave in Fig. 3(a) one can write the propagation factor as,
\begin{eqnarray}
\label{33}
\frac{V''_{0}}{V''_{1}}=\frac{I''_{0}}{I''_{1}}=1-\omega^2 LC/2-j\sqrt{\omega^2LC}\sqrt{1 - \omega^2 LC/4}.
\end{eqnarray}
Equation (\ref{31}) can be rewritten as,
\begin{eqnarray}
\label{31a}
\frac{V'_{0}}{V'_{1}}=\frac{I'_{0}}{I'_{1}}=1-\omega^2 LC/2+j\sqrt{\omega^2LC}\sqrt{1 - \omega^2 LC/4},
\end{eqnarray}

From Eqs. (\ref{33}) and (\ref{31a}) we get for the voltage $V_0$ and current $I_0$ as the superposition of the incident and reflected waves,
\begin{eqnarray}
\label{36}
V_0=V'_{0}+V''_{0}=2 V'_{1}(1-\omega^2 LC/2)\\
\label{37}
I_0=I'_{0}+I''_{0}=2 I'_{1} j\sqrt{\omega^2LC}\sqrt{1 - \omega^2 LC/4}
\end{eqnarray}
Therefore we get the input impedance $Z_i$ as,
\begin{eqnarray}
\label{38}
Z_i=V_0/I_0=\frac{V'_{1}}{I'_{1}}\frac{1-\omega^2 LC/2}{j\sqrt{\omega^2LC}\sqrt{1 - \omega^2 LC/4}}
\end{eqnarray}
Using $Z_0=\sqrt{L/C -\omega^2 L^2/4}=\sqrt{L/C}\sqrt{1 -\omega^2 LC/4}$ , we get
$Z_i=j\omega L/2 +1/(j\omega C)$ , which of course is the expected result Eq. (\ref{30}). The input impedance is {\em imaginary} 
for all driving frequencies.
\section*{Acknowledgements}
I thank Prof. S. C. Dutta Roy of IIT Delhi for his comments and suggestions on the manuscript.

\end{document}